\documentclass[12pt,onecolumn]{emulateapj}

 \makeatother

\def\vol#1  {{{#1}{\rm,}\ }}

\newcount\refno
\newcount\rfno
\def\eq{$^{\the\refno\ }$\advance\refno by 1}
\def\ad{\advance\rfno by 1}

\def\clock{\count0=\time \divide\count0 by 60
     \count1=\count0 \multiply\count1 by -60 \advance\count1 by \time
     \number\count0:\ifnum\count1<10{0\number\count1}\else\number\count1\fi}

\def\myputfigure#1#2#3#4#5%
{\hskip0.03\textwidth\vskip#5pt
\makebox[0pt]{\hskip#2in
\includegraphics[width=#3\textwidth]{#1}}\vskip#4pt\hfill}

\def\Gcm2{\rm G~cm^2}

\def\beq{\begin{equation}}
\def\eeq{\end{equation}}
\def\bea{\begin{eqnarray}}
\def\eea{\end{eqnarray}}

\def \date         {\ifcase\month \message{zero} \or
                    January \or February \or March \or April \or May \or June
                    \or July \or
                    August \or September \or October \or November \or
                    December \fi
                    \space\number\day, \number\year}

\begin{document}

\title{A pure hydrodynamic instability in shear flows and its application to astrophysical accretion disks}
\author{Sujit Kumar Nath\altaffilmark{1,2} and Banibrata Mukhopadhyay\altaffilmark{1,3}}
\altaffiltext{1}{Department of Physics, Indian Institute of Science, Bangalore-560012}
\altaffiltext{2}{sujitkumar@physics.iisc.ernet.in}
\altaffiltext{3}{bm@physics.iisc.ernet.in}

\begin{abstract}
We provide the possible resolution for the century old problem of hydrodynamic shear flows,
which are apparently stable in linear analysis but shown to be turbulent in 
astrophysically observed data and experiments.
This mismatch is noticed in a variety of systems, from 
laboratory to astrophysical flows. 
There are so many uncountable attempts made so far to resolve
this mismatch, beginning with the early work of
Kelvin, Rayleigh, and Reynolds towards the end of the nineteenth
century.
Here we show that the presence of stochastic noise, whose inevitable presence should not be neglected
in the stability analysis of shear flows, leads to pure hydrodynamic linear instability therein. 
This explains the origin of turbulence, which has been observed/interpreted
in astrophysical accretion disks, laboratory experiments and direct numerical simulations.
This is, to the best of our knowledge, the first solution to the 
long standing problem of hydrodynamic instability of Rayleigh stable flows.

\end{abstract}


\section{Introduction}\label{intro}

The astrophysically ubiquitous Keplerian accretion disks should be unstable and 
turbulent in order to explain observed data, but are remarkably Rayleigh stable.
They are found in 
active galactic nuclei (AGNs), around a compact object in binary systems, around newly formed 
stars etc. \citep[see, e.g.,][]{pringle}. 
The main puzzle of accreting material in disks is its inadequacy of 
molecular viscosity to transport them towards the central object. 
Thus the idea of turbulence and, hence, turbulent viscosity has been proposed.
Similar issue is there in certain shear flows, e.g. plane Couette flow, which are shown 
to be linearly stable for any Reynolds number ($Re$) but in laboratory could be turbulent for 
$Re$ as low as $350$. 
Therefore, linear perturbation cannot induce the turbulent viscosity to transport matter inwards and angular momentum outwards, 
in the Keplerian disks. Note that the
issue of linear instability of the Couette-Taylor flow (when accretion disks are the subset of it) is a 
century old problem.

Although in the presence of vertical shear and/or stratification, 
Keplerian flow may reveal Rayleigh-Taylor type instability
(e.g. \citealt{nelson,stoll,barker,lin,ric,um,stoll2}),
convective overstability (\citealt{klar,lyra}) and the Zombie 
Vortex Instability (\citealt{marcus2,marcus1}), we intend here to solve 
the classic century old problem of the origin of linear instability with the 
exponential growth of perturbation in purely hydrodynamical Rayleigh-stable 
flows with only radial shear. The convective overstability does not 
correspond to an indefinitely growing mode and it has some saturation
(\citealt{latter}). In addition, the Zombie 
Vortex Instability is not sufficient to transport 
angular momentum significantly in a small domain of study. In fact,
all of them could exhibit only smaller Shakura-Sunyaev viscosity parameter (\citealt{ss73}) $\alpha_{ss}<10^{-3}$ than 
that generally required to explain observation. The robustness of our work
is that, it can explain the turbulent behavior of any kind of
Rayleigh-stable shear flows, starting from laboratory to astrophysical
flows. While many realistic non-magnetized and Keplerian flows could be 
stratified in both the vertical and radial directions of the disks, 
it is perhaps impossible to prove that all the non-magnetized accretion
disks have significant amount of vertical shear and/or stratification 
to sustain the above
mentioned instabilities. Note that indeed many accretion disks are 
geometrically thin. Moreover, the laboratory Taylor-Couette flows
have no vertical shear and/or stratification.

In 1991, with the application of Magnetorotational Instability 
 \citep[MRI;][]{velikhov,chandra} to Keplerian disks, \cite{bh} showed
that initial weak magnetic field can lead to the perturbations
growing exponentially. 
Within a few rotation times, such exponential growth could reveal the onset of turbulence. 
However, for charge neutral flows
MRI should not work. Note also that for flows having strong magnetic fields, where the magnetic field 
is tightly coupled with the flow, MRI is not expected to work 
(e.g. \citealt{tran}). 
It is a long standing controversy \citep[see, e.g.,][]{dau,zahn,gu,kim,rud,klar1,yecko,amn,man,
dub1,dub2,mk,bmraha,mc13},
whether the matter in Rayleigh stable astrophysical disks is stable or unstable. The answer has profound significance for our 
understanding of how stars and planets form. It is argued, however, that some types of Rayleigh stable 
flows certainly can be destabilized \citep{stattur,avilasci,balbusnature,fulltur}.
Based on `shearing sheet' approximation, without \citep{balbusetal96,hawleyetal99}
and with \citep{ll05} explicit viscosity, some authors attempted to tackle the issue of turbulence
in hot accretion disks. However, other authors argued for 
limitations in this work \citep{pumir96,fromang-papaloizou07}. 
Based on the simulations including explicit viscosity,
the authors could achieve $Re\approx 4\times 10^4$ and concluded that Keplerian like 
flows could exhibit very weak turbulence in the absence of magnetic field. 
Nevertheless, the recent experimental 
results by \cite{paoletti} clearly argued for the significant level of transport 
from hydrodynamics alone. Moreover, the results from direct numerical simulations  
\citep{avila} and exploration of transient amplification, in otherwise linearly stable flows, 
with and without noise \citep[e.g.][]{man,tref,berk}
also argued for (plausible) hydrodynamic instability and turbulence at low $Re$. 
Interestingly, accretion disks have huge $Re$ ($\gtrsim10^{15}$) \citep{bmplb}, prompting to the 
belief that they are hydrodynamically unstable.

We show here that linearly perturbed apparently Rayleigh stable flows driven stochastically can be
made unstable even in the absence of any magnetic field.
We also argue, why stochastic noise is inevitable 
in such flows. They exist in the flows under consideration inherently. 
We develop our theory following the seminal concept 
based on fluctuating hydrodynamics of randomly stirred fluid,
pioneered by \cite{nelson} and \cite{dm}, which, however, was never applied 
in the context of accretion flows or other shear flows. This work provides a new path of 
linear hydrodynamic instability of shear flows, which will have vast applications from 
accretion disks to laboratory flows, for the first time.

The plan of the paper is the following. In the next section, we introduce equations
describing the system under consideration. Then \S 3 describes the evolution
of various perturbations in stochastically driven hydrodynamic flows. 
Subsequently, we discuss the relevance of white noise in the context of shear flows in \S 4.
Finally we summarize with conclusions in \S 5.
In appendix, we demonstrate in detail the generation of white noise from random walk, 
particularly in the present context.

\section{Equations describing perturbed rotating shear flows in the presence of noise}

The linearized Navier-Stokes equation in the presence of background plane shear $(0,-x,0)$ 
and angular velocity $\Omega\propto r^{-q}$, 
when $r$ being the distance from the center of the system,
in a small section approximated as incompressible flow with $-1/2\le x\le 1/2$, 
has already been established \citep{mc13}. 
Here, any length is expressed in units of the size $L$ of the system in the $x-$direction, the time in
units of $\Omega^{-1}$,
the velocity in $q\Omega L$ ($1\le q<2$), and other variables are expressed accordingly 
(see, e.g., \citealt{man,bmraha,mc13}, for detailed description of the
choice of coordinate in a small section). Hence, in dimensionless units, the 
linearized Navier-Stokes equation and 
continuity equation (for an incompressible flow) 
can be recasted into the well-known Orr-Sommerfeld and Squire equations, but in the presence of stochastic noise and 
Coriolis force \citep{mc13}, given by 

\begin{equation}
\left(\frac{\partial}{\partial t}-x\frac{\partial}{\partial y}\right)\nabla^2 u
+\frac{2}{q}\frac{\partial \zeta}{\partial z}=\frac{1}{Re}\nabla^4 u+\eta_1(x,t),
\label{hydroorrv}
\end{equation}
\begin{equation}
\left(\frac{\partial}{\partial t}-x\frac{\partial}{\partial y}\right)\zeta
+\frac{\partial u}{\partial z}\left(1-\frac{2}{q}\right)=\frac{1}{Re}\nabla^2 \zeta +
\eta_2(x,t),
\label{hydroorrzeta}
\end{equation}
where $\eta_{1,2}$ are the components of noise arising in the linearized system due 
to stochastic forcing such that $<\eta_i(\vec x,t) \eta_j(\vec x',t')>=D_i(\vec x)\:\delta^3(\vec x-\vec x')\:\delta(t-t')\:\delta_{ij}$ \citep{nelson}, 
where $D_i(\vec x)$ is a constant for white noise and $i,j=1,2$; $u$ is the $x$-component of velocity perturbation vector and 
$\zeta$ the $x$-component of vorticity perturbation vector.

Now, we can resort to a Fourier series expansion of
$u$, $\zeta$ and $\eta_i$ as
\begin{equation}
A(\vec{x},t)=\int\tilde{A}_{\vec{k},\omega}\,e^{i(\vec{k}.\vec{x}-\omega t)}d^3k\,d\omega,\\
\label{fourier}
\end{equation}
where $A$ can be any one of $u$, $\zeta$ and $\eta_i$; $\vec{k}$ and $\omega$ are the wavevector and frequency respectively 
in the Fourier space such that $\vec{k}=(k_x,k_y,k_z)$ and $|\vec{k}|$=$k$.


\section{Evolution of perturbation in stochastically driven hydrodynamic accretion flows}\label{plots}
%
Writing down equations (\ref{hydroorrv}) and (\ref{hydroorrzeta}) in Fourier space by using equation (\ref{fourier}), 
and taking ensemble average, 
we obtain the equations involving the evolution of mean values of 
perturbations in the presence of noise as
\begin{equation}
2\pi k_yk^2\frac{\partial\tilde{<u>}_{\vec{k},\omega}}{\partial k_x}=\left(i\omega k^2-4\pi k_xk_y-\frac{k^4}{R_e}\right)
\tilde{<u>}_{\vec{k},\omega}
+\frac{2ik_z}{q}\tilde{<\zeta>}_{\vec{k},\omega}-m\delta(\vec{k})\delta(\omega),
\label{hydrofourierumean}
\end{equation}

\begin{equation}
2\pi k_y\frac{\partial\tilde{<\zeta>}_{\vec{k},\omega}}{\partial k_x}=-ik_z\left(1-\frac{2}{q}\right)\tilde{<u>}_{\vec{k},\omega} 
+\left(i\omega-\frac{k^2}{R_e}\right)\tilde{<\zeta>}_{\vec{k},\omega}+m\delta(\vec{k})\delta(\omega),
\label{hydrofourierzetamean}
\end{equation}
where Fourier transformations of $\eta_{1,2}$ are
basically $\delta(\vec{k})\delta(\omega)$ multiplied with a random number and on ensemble average it
appears to be a constant $m$ which is the mean value of the white noise (we get $m=0$ and $m\neq 0$ when the drift coefficient 
of the Brownian motion or Wiener process corresponding to the white noise is zero and nonzero 
respectively, see Appendix for details), and 
$<u>_{\vec{k},\omega}$, $<\zeta>_{\vec{k},\omega}$ are the Fourier transforms of $<u>$ and $<\zeta>$ which are the mean or 
the ensemble averaged values of $u$ and $\zeta$ respectively.

\subsection{Evolution of vertical perturbations}

Now let us take the trial solutions, 
$<u>,<\zeta>=u_0,\zeta_0\exp\ i(\vec{\alpha}.\vec{x}-\beta t)$, where 
$u_0,\zeta_0$ are the constant, in general complex, amplitudes of perturbation and $\vec{\alpha}=(0,0,\alpha)$, is a vertical wavevector (one should not confuse this $\alpha$ with the Shakura-Sunyaev viscosity parameter). 
Vertical wave vector is chosen since it will be unaffected by shear \citep{bh}. This
gives $\tilde{<u>}_{\vec{k},\omega},\tilde{<\zeta>}_{\vec{k},\omega}= 
u_0,\zeta_0\ \delta(\vec{\alpha}-\vec{k})\delta(\beta-\omega)$ 
(using equation (\ref{fourier})). 
Substituting these trial solutions in equations (\ref{hydrofourierumean}) and (\ref{hydrofourierzetamean}), 
integrating with respect to $\vec{k}$ and $\omega$ we obtain 

\begin{equation}
\left(i\beta \alpha^2-\frac{\alpha^4}{Re}\right)u_0+\frac{2i\alpha}{q}\zeta_0-m=0,
\label{hydroudispersion}
\end{equation}

\begin{equation}
-i\alpha\left(1-\frac{2}{q}\right)u_0+\left(i\beta-\frac{\alpha^2}{Re}\right)\zeta_0+m=0.
\label{hydrozetadispersion}
\end{equation}

\subsubsection{Case I}

Now eliminating $m$ and assuming $\zeta_0=i u_0$ we obtain the dispersion relation
\begin{equation}
\left(i\beta \alpha^2-\frac{\alpha^4}{Re}\right)-\frac{2\alpha}{q}= 
i\alpha\left(1-\frac{2}{q}\right)+\left(\beta+\frac{i\alpha^2}{Re}\right).
\label{vertmeli}
\end{equation}
If we find any pair of $\alpha$ and $\beta$
satisfying equation (\ref{vertmeli}) 
for which the imaginary part of $\beta$
positive, then we can say that the mean value of perturbation is unstable.
Equation (\ref{vertmeli}) is the hydrodynamic counter part of the dispersion 
relation obtained due to
MRI \citep{bh}, leading to the avenue of pure hydrodynamic instability. 
For $m=0$, from equations (\ref{hydroudispersion}) and 
(\ref{hydrozetadispersion}), either $u_0$ and $\zeta_0$ both turn out to be zero or there is 
no instability for non-trivial $u_0$ and $\zeta_0$.
Overall, $m=0$ gives rise to stable solutions like the zero magnetic field for MRI.

\begin{figure}[tbp]
\centering
\includegraphics[scale=1.2]{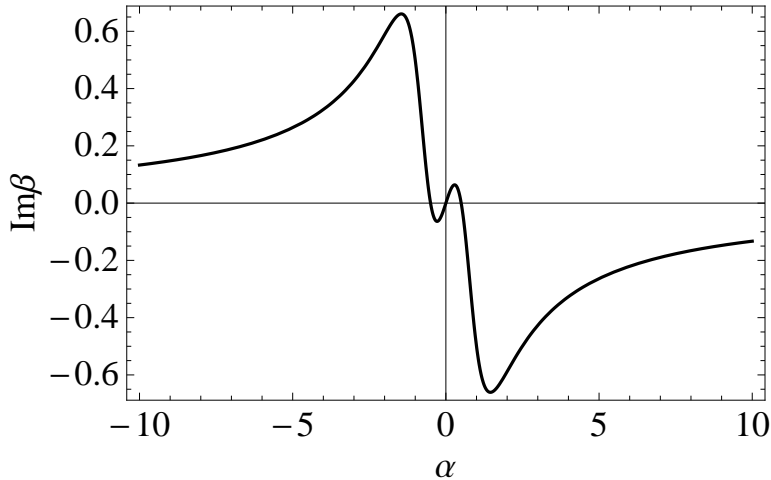}
\caption{Relationship between $\alpha$ and the imaginary part of $\beta$, for
vertical perturbation in Case I with $Re=10^7$. We consider $q = 1.5$, however we obtain almost 
the same results for other admissible values of $q$. }
\label{vertm}
\end{figure}

Figure \ref{vertm} shows the ranges of $\alpha$ giving rise to linear instability.
It is easy to understand that similar results could be obtained with the choice of unequal
ensemble averages of white noise in equations (\ref{hydroudispersion}) and 
(\ref{hydrozetadispersion}) and a more general phase difference between $\zeta_0$ and $u_0$.
\vskip1.5cm

\subsubsection{Case II}

Now for a given $u_0$ and $m$, after eliminating $\zeta_0$ from equations 
(\ref{hydroudispersion}) and (\ref{hydrozetadispersion}), 
we obtain a dispersion relation between $\alpha$ and $\beta$ as 
%

\begin{equation}
\alpha^2\beta^2+i\beta\left(\frac{2\alpha^4}{Re}+\frac{m}{u_0}\right) 
+\left(\frac{2im\alpha}{qu_0}-\frac{4\alpha^2}{q^2}+\frac{2\alpha^2}{q}-\frac{m\alpha^2}{u_0Re}-\frac{\alpha^6}{{Re}^2}\right)=0,
\label{dispersion}
\end{equation}
which is second order in $\beta$ and hence has two roots $\beta_1$ and $\beta_2$. 
If we find any pair of $\alpha$ and $\beta$ 
for which the imaginary part of $\beta$ 
positive, then we can say that the mean value of perturbation is unstable.
For $m=0$ in equation (\ref{dispersion}), there is no instability,
like the zero magnetic field for MRI.
%
%
\begin{figure}[tbp]
\centering
  \begin{tabular}{@{}cc@{}}
    \includegraphics[width=.5\textwidth,height=0.25\textheight]{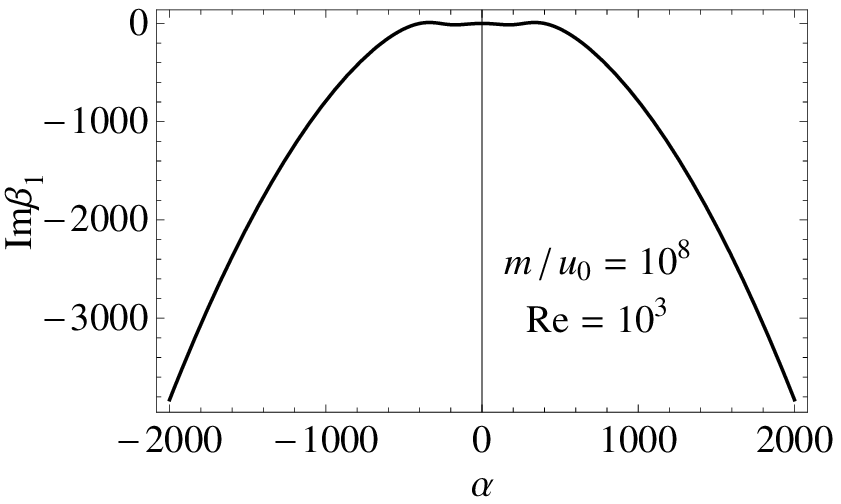} 
    \includegraphics[width=.5\textwidth,height=0.25\textheight]{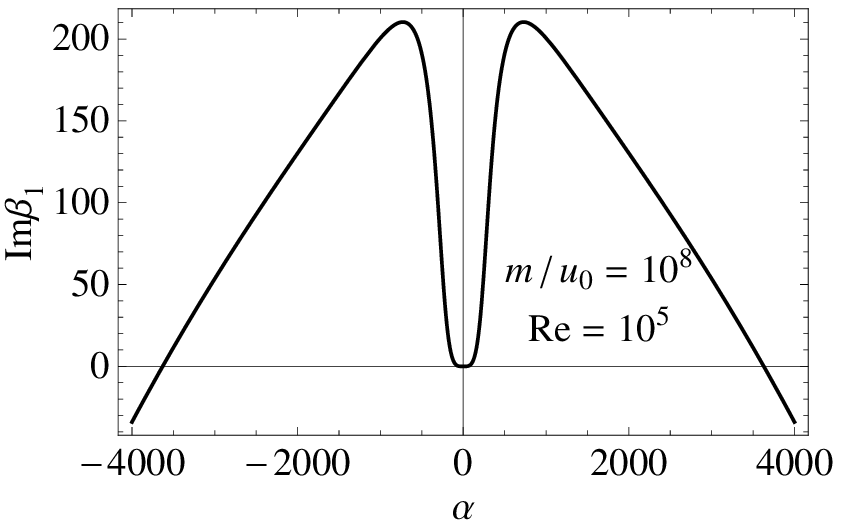} \\ \\ \\ \\ 
    \includegraphics[width=.5\textwidth,height=0.25\textheight]{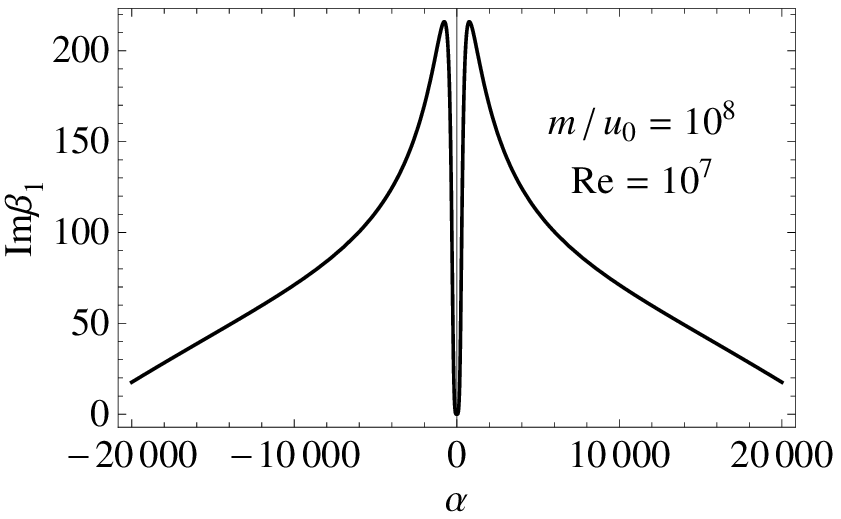} 
    \includegraphics[width=.5\textwidth,height=0.25\textheight]{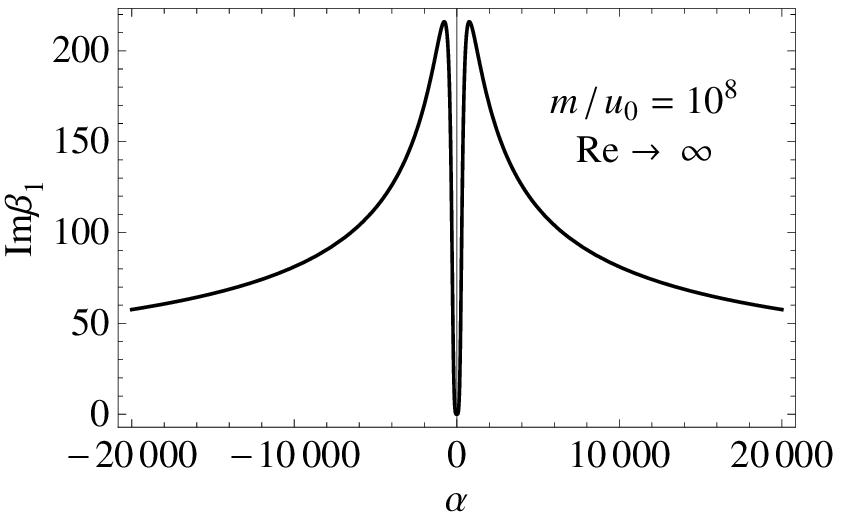} 
    \end{tabular}
  \caption{
Relationship between $\alpha$ and the imaginary part of one of the solutions of $\beta$,
for vertical perturbation in Case II.
We consider $q=1.5$, however we obtain 
almost the same results for other admissible values of $q$. Other 
solution of $\beta$ is stable.}
\label{unstable}
\end{figure}
\vskip1.5cm
%

In Fig. \ref{unstable}, for $m/u_0=10^8$ and different values of $Re$ above a certain value, 
we show that for Keplerian flows, there are 
modes for which the mean values of perturbation are unstable. 
If the amplitude of perturbations decreases, the value of $m/u_0$ increases for any fixed 
nonzero $m$, leading to a larger range of $\alpha$ for instability. 
However for $m=0$, i.e. for the white noise with zero mean 
(which also corresponds to the hydrodynamic accretion flows without any noise), we obtain no such unstable modes. 
While modes are stable for smaller $Re$, with the increase of $Re$ they become unstable and range of $\alpha$ 
giving rise to instability increases with increasing $Re$ and for 
$Re\rightarrow\infty$ unstable modes arise all the way upto $|\alpha|\rightarrow\infty$.
%
%
\begin{figure}[tbp]
\centering
  \begin{tabular}{@{}cc@{}}
    \includegraphics[width=.5\textwidth]{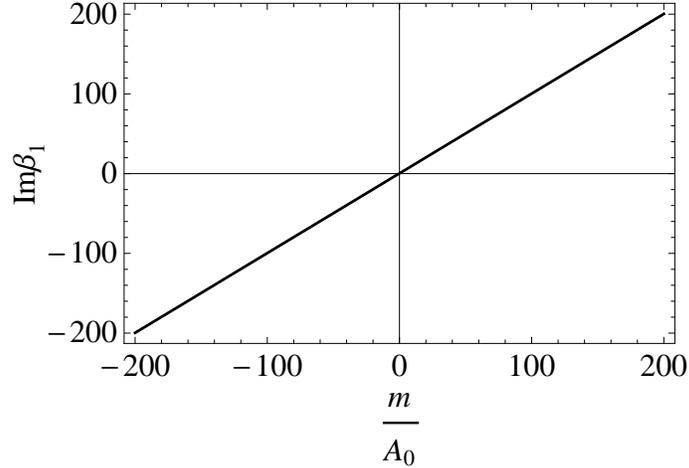} \\ \\ 
    \end{tabular}
  \caption{
Relationship between $m/A_0$ and the imaginary part of one of the solutions of $\beta$, for 
vertical perturbation in Case III.
We consider $q=1.5$, however we obtain almost the same results for other admissible values of $q$.}
\label{unstablem}
\end{figure}


\subsubsection{Case III}

Now we assume, for simplicity and without loss 
of much generality, $u_0=\zeta_0=A_0$. 
Then expressing $\alpha$ in terms of $m/A_0$ from equations 
(\ref{hydroudispersion}) and (\ref{hydrozetadispersion}), by means 
of a cubic equation, 
given by
\begin{eqnarray}
i\left(1-\frac{2}{q}\right)\alpha^3-\frac{m}{A_0}\alpha^2+\frac{2i}{q}\alpha-\frac{m}{A_0}=0
\end{eqnarray}
and supplemented by equation (\ref{hydrozetadispersion}), we obtain three roots of $\beta$.
Figure \ref{unstablem} shows that the first 
solution of $\beta$ ($\beta_1$) exhibits unstable modes for any $m/A_0>0$ 
(however small the magnitude be), which is also independent of $Re$ (however
$\beta_2$ and $\beta_3$ need not be $Re$ independent). 
Therefore, if we have any stochastic forcing with arbitrarily small but 
fixed nonzero value of $m$ (drift coefficient), we always have unstable mean perturbation modes 
since $A_0$ can be made arbitrarily small.

\subsubsection{Plane Couette flow and negative $m$}

\begin{figure}[tbp]
\centering
\includegraphics[scale=1.2]{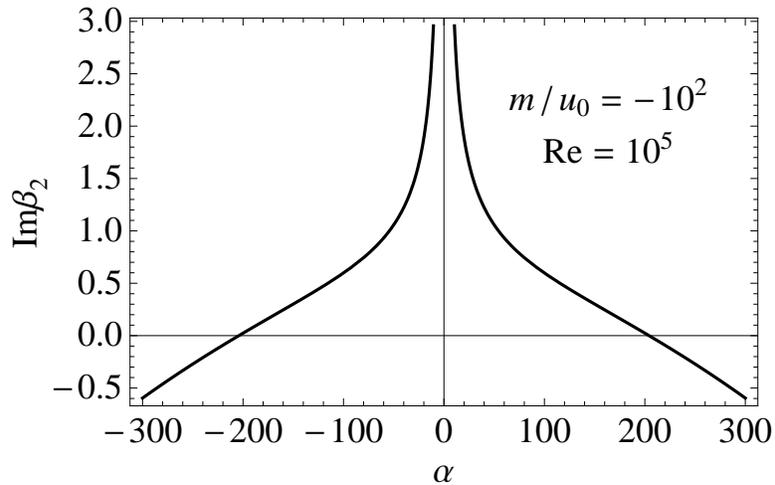}
\caption{Relationship between $\alpha$ and imaginary part of one of the solutions of $\beta$,
for vertical perturbation in Case II, where $q=1.5$. It shows the unstable modes for 
negative $m$ (drift velocity).}
\label{negm}
\end{figure}

Figures \ref{unstable} and \ref{unstablem} demonstrate instability for positive $m$ and real $\alpha$.
However, negative $m$ with the appropriate choice of $\alpha$ (real or complex) could also lead to instability
for Keplerian flows (the same is true for positive $m$ and complex $\alpha$). For plane Couette flow, however,
in order to demonstrate instability, either $m$ has to be negative with real $\alpha$ or $\alpha$ has to be
complex with positive $m$. From equation (\ref{dispersion}), for $q\rightarrow\infty$ (i.e. plane Couette flow), we 
obtain the corresponding two possible dispersion relations as
\begin{eqnarray} 
\beta=-i\frac{\alpha^2}{Re},~~-i\frac{Re~m/u_0+\alpha^4}{Re~\alpha^2}.
\label{cou}
\end{eqnarray} 
The second solution will
lead to the instability for a negative $m$ satisfying $|m|/u_0>\alpha^4/Re$.
Note that for real $\alpha$, there are always appropriate values of $m$ leading to instability in both Keplerian and 
plane Couette flows. For negative $m$, the Keplerian flows remain unstable upto $|\alpha|\rightarrow 0$,
as shown in Fig. \ref{negm}, unlike the positive $m$ cases.
%
%
\subsection{Evolution of perturbations with spherical modes} 
In this section we show that there are other perturbation modes also which are linearly unstable. Here 
we show this for spherical modes as an example. 
However, such perturbation modes might be taken only under the 
assumption that they do not get distorted much due to shear, which may 
not be completely correct (\citealt{amn,man,tran}).
Writing down equations (\ref{hydrofourierumean}) and 
(\ref{hydrofourierzetamean}) for spherical wave (i.e. $k_x=k_y=k_z=k/\sqrt{3}$), 
we obtain the equations involving the evolution of mean values of perturbations in the presence of noise as 
\begin{eqnarray} 
2\pi k^3\frac{\partial\tilde{<u>}_{\vec{k},\omega}}{\partial k}=\left(i\omega k^2-\frac{4\pi k^2}{3}
-\frac{k^4}{Re}\right)\tilde{<u>}_{\vec{k},\omega}, 
+\frac{2ik}{\sqrt{3}q}\tilde{<\zeta>}_{\vec{k},\omega}+m\delta(\vec{k})\delta(\omega), 
\label{hydrofourierumeansp}
\end{eqnarray}

\begin{eqnarray} 
2\pi k\frac{\partial\tilde{<\zeta>}_{\vec{k},\omega}}{\partial k}= 
-\frac{ik}{\sqrt{3}}\left(1-\frac{2}{q}\right)\tilde{<u>}_{\vec{k},\omega} 
+\left(i\omega-\frac{k^2}{Re}\right)\tilde{<\zeta>}_{\vec{k},\omega}+m\delta(\vec{k})\delta(\omega). 
\label{hydrofourierzetameansp}
\end{eqnarray}
Substituting the trial solutions for $<u>$ and $<\zeta>$, as described above 
equation (\ref{hydroudispersion}) but replacing vertical $\alpha$ by 
spherical $\alpha$,  
in equations (\ref{hydrofourierumeansp}) and (\ref{hydrofourierzetameansp}), and integrating with respect to 
$\vec{k}$ and $\omega$ we obtain 
\begin{eqnarray}
6\pi \alpha^2u_0=\left(i\beta \alpha^2-\frac{4\pi \alpha^2}{3}-\frac{\alpha^4}{Re}\right)u_0 
+\frac{2i\alpha}{\sqrt{3}q}\zeta_0+m,
\label{hydroudispersionsp}
\end{eqnarray}

\begin{eqnarray}
2\pi\zeta_0=-\frac{i\alpha}{\sqrt{3}}\left(1-\frac{2}{q}\right)u_0 
+\left(i\beta-\frac{\alpha^2}{Re}\right)\zeta_0+m.
\label{hydrozetadispersionsp}
\end{eqnarray} 

\subsubsection{Case I}

Now eliminating $m$ from equations (\ref{hydroudispersionsp}) and (\ref{hydrozetadispersionsp}) and assuming $\zeta_0=i u_0$ we obtain the dispersion relation
\begin{eqnarray}
6\pi \alpha^2-\left(i\beta \alpha^2-\frac{4\pi \alpha^2}{3}-\frac{\alpha^4}{Re}\right)
+\frac{2\alpha}{\sqrt{3}q}=2\pi i+\frac{i\alpha}{\sqrt{3}}\left(1-\frac{2}{q}\right)
+\left(\beta+\frac{i\alpha^2}{Re}\right)
\label{spmeli}
\end{eqnarray}
\begin{figure}[tbp]
\centering
\includegraphics[angle=0,width=8.5cm]{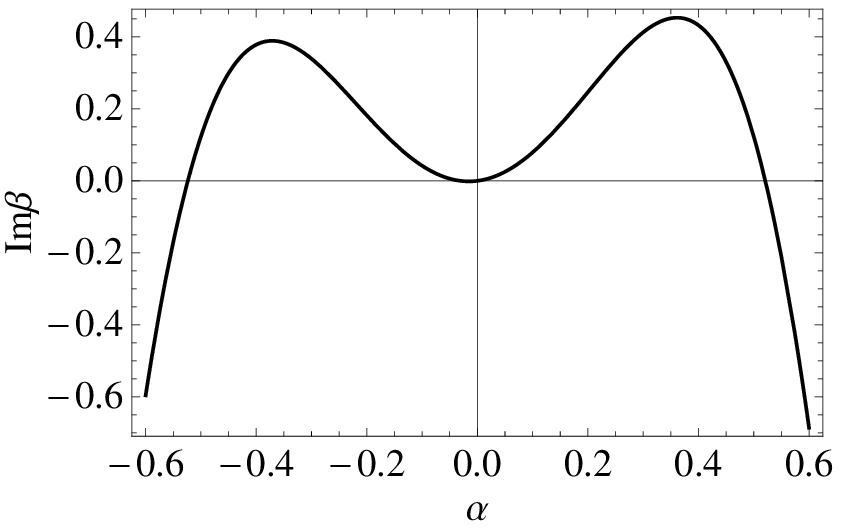}
\caption{Relationship between $\alpha$ and the imaginary part of $\beta$, for
spherical perturbation in Case I with $Re=10^7$. We consider $q = 1.5$, however we obtain almost
the same results for other admissible values of $q$.
}
\label{spm}
\end{figure}
\vskip1.5cm

Figure \ref{spm} shows the ranges of $\alpha$ giving rise to linear instability.
It is easy to understand that similar results could be obtained with the choice of unequal
ensemble averages of white noise in equations (\ref{hydroudispersionsp}) and
(\ref{hydrozetadispersionsp}) and a more general phase difference between $\zeta_0$ and $u_0$.

\subsubsection{Case II}

For a given $u_0$ and $m$, after eliminating $\zeta_0$ from equations (\ref{hydroudispersionsp}) and 
(\ref{hydrozetadispersionsp}), we obtain a dispersion relation between $\alpha$ and $\beta$ for spherical perturbation as 
%
%
\begin{eqnarray} 
\nonumber
\alpha^2\beta^2+i\beta\left(\frac{2\alpha^4}{Re}+\frac{28\pi\alpha^2}{3}+\frac{m}{u_0}\right)+\frac{2}{3q}
\left(1-\frac{2}{q}\right)\alpha^2 \\ 
-\left(2\pi+\frac{\alpha^2}{Re}\right)\left(\frac{22\pi\alpha^2}{3}+\frac{\alpha^4}{Re}\right) 
-\left(2\pi+\frac{\alpha^2}{Re}-\frac{2i\alpha}{\sqrt{3}q}\right)\frac{m}{u_0}=0.
\label{dispersionsp}
\end{eqnarray}
\begin{figure}[tbp]
\centering
  \begin{tabular}{@{}cc@{}}
    \includegraphics[width=.5\textwidth,height=0.25\textheight]{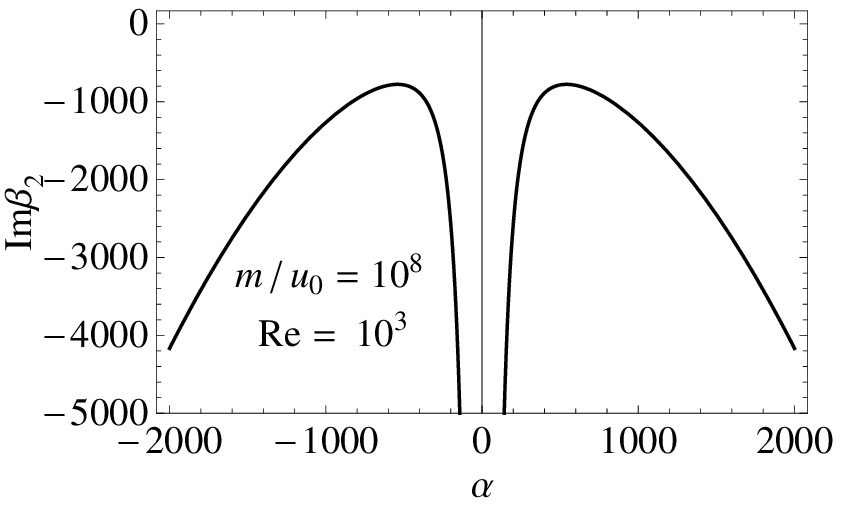} 
    \includegraphics[width=.5\textwidth,height=0.25\textheight]{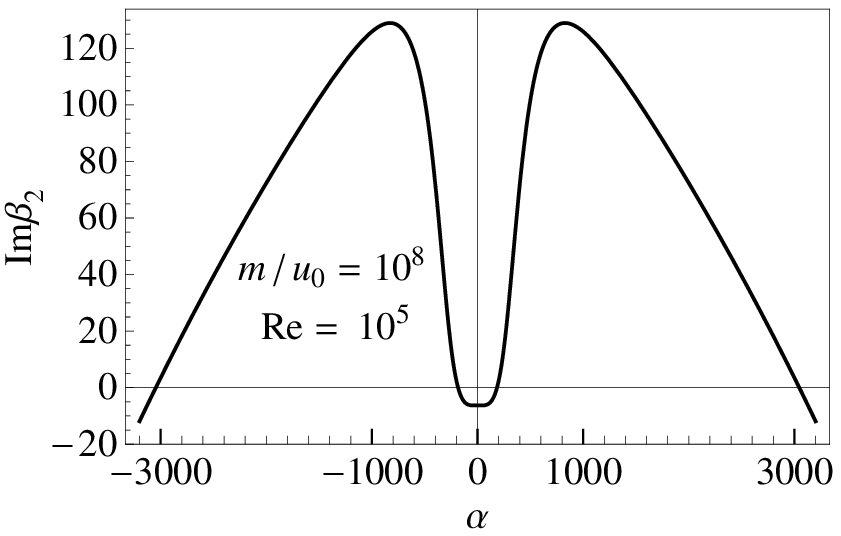} \\ \\ \\ \\ 
    \includegraphics[width=.5\textwidth,height=0.25\textheight]{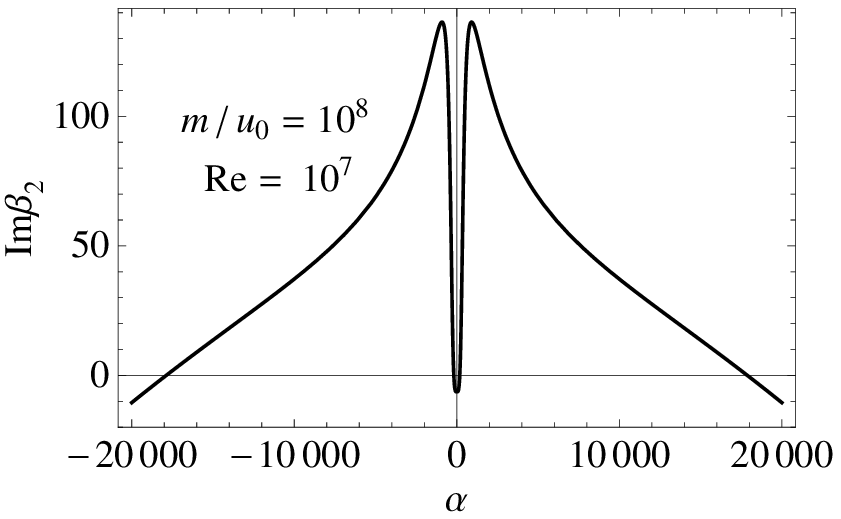} 
    \includegraphics[width=.5\textwidth,height=0.25\textheight]{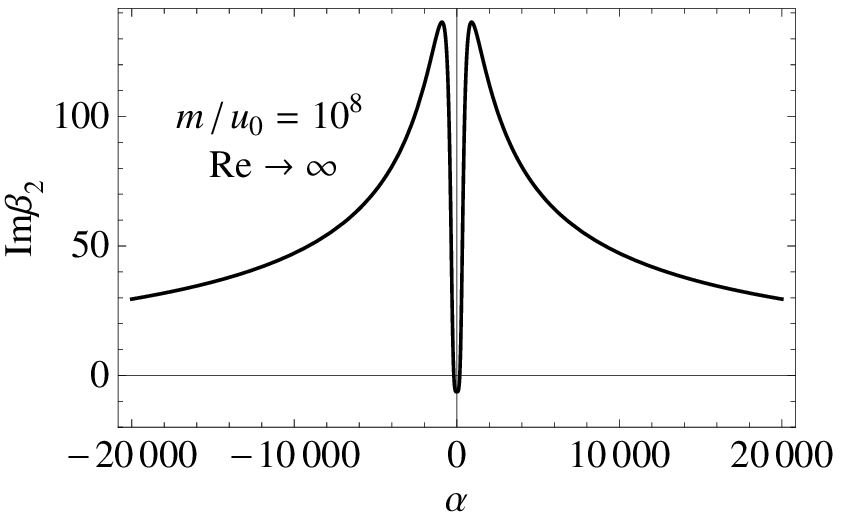} 
    \end{tabular}
  \caption{
Relationship between $\alpha$ and the imaginary part of one of the 
solutions of $\beta$, for spherical perturbation in Case II.
We consider $q=1.5$, however we obtain 
almost the same results for other admissible values of 
$q$. Other solution of $\beta$ is stable.}
\label{unstablesp}
\end{figure}
In Fig. \ref{unstablesp}, for $m/u_0=10^8$ and different values of $Re$, we show that for Keplerian flows, there are 
several spherical modes, for which the mean values of perturbations grow exponentially, just like Fig. \ref{unstable} 
shows the same for vertical perturbations. 
\begin{figure}[tbp]
\centering
  \begin{tabular}{@{}cc@{}}
    \includegraphics[width=.5\textwidth]{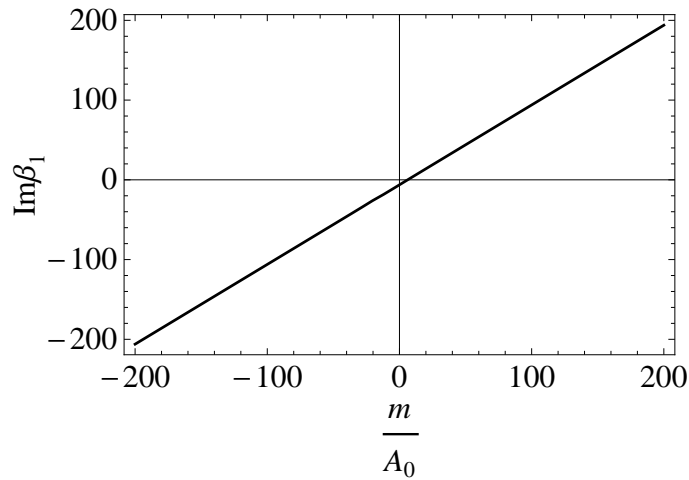} 
    \end{tabular}
  \caption{
Relationship between $m/A_0$ and the imaginary part of the solutions
of $\beta$, for spherical perturbation in Case III. We consider $q=1.5$, however 
we obtain almost the 
same results for other admissible values of $q$.}
\label{unstablemsp}
\end{figure} 

\subsubsection{Case III} 
 
In Fig. \ref{unstablemsp}, we show how spherical perturbation modes in Keplerian flows vary with $m/A_0$. This is very similar to 
as shown in Fig. \ref{unstablem} for vertical perturbations, except that the modes
are stable for a very small but non-zero $m/A_0$, while for vertical 
perturbation the modes remain unstable for $m/A_0\rightarrow 0$.  

\subsubsection{Plane Couette flow and negative $m$}

For plane Couette flows, making $q\rightarrow\infty$ in 
equation (\ref{dispersionsp}), we obtain the corresponding dispersion relation as 
\begin{eqnarray} 
\beta=-i\frac{2\pi Re+\alpha^2}{Re},~~-i\frac{3Rem/u_0+22\pi Re\alpha^2+3\alpha^4}{3Re\alpha^2}.
\label{cousp}
\end{eqnarray} 
While the first root always corresponds to the stable mode for a real $\alpha$, the second one will
lead to the unstable solution for a negative $m$ satisfying $|m|/u_0>22\pi\alpha^2/3+\alpha^4/Re$. 
\begin{figure}[tbp]
\centering
\includegraphics[angle=0,width=8.5cm]{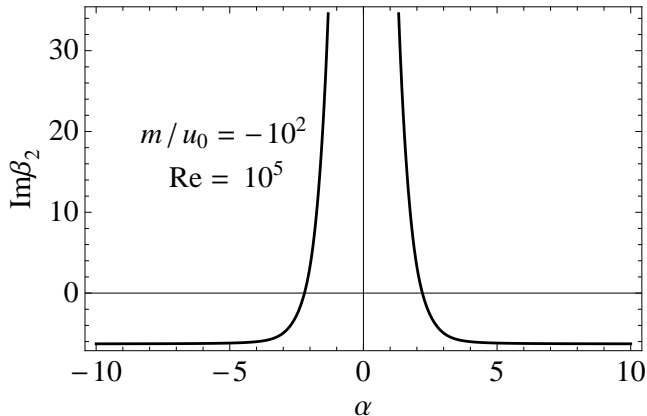}
\caption{Relationship between $\alpha$ and imaginary part of one of the
solutions of $\beta$, for spherical perturbation in Case II, where $q=1.5$. 
It shows the unstable modes for negative $m$ (drift velocity).}
\label{negmsp}
\end{figure}
Figure \ref{negmsp} shows that for spherical perturbations, the Keplerian flows remain unstable 
upto $|\alpha|\rightarrow 0$, as shown in Fig. \ref{negm} for vertical perturbation cases. 


\section{Relevance of white noise in the context of shear flows}\label{noise}

Now we shall discuss that how relevant and how likely the white 
noise is to be present in shear flows. 
The Rayleigh stable flows under consideration have a background shear 
profile, with some molecular viscosity however small that may be, 
and hence some drag (e.g., in protoplanetary disks, it could be due to the 
drag between gas and solid particles). For plane Couette flow, such shear is driven in the fluids
by moving the boundary walls by externally applied force. If the external force (cause) 
is switched off, the shearing motion (effect) dies out. Similarly, in accretion disks,
the central gravitational force plays the role of driving force (cause) producing 
differential velocity (shear) in the flow. Hence, by 
fluctuation-dissipation theorem of statistical mechanics 
\citep[see, e.g.,][]{physica,prlf}, there must be some thermal 
fluctuations in such flows, with some temperature however low be, and that cause 
the fluid particles to have Brownian motion. Therefore the time variation (derivative) of this Brownian motion, which is defined as white 
noise, plays the role of extra stochastic forcing term in the Orr-Sommerfeld equations (equations (\ref{hydroorrv}), 
(\ref{hydroorrzeta})) which are present generically, in particular when perturbation is considered.

Now, due to the presence of background shear in some preferential direction, it is very likely for the 
fluid particles to have Brownian motion with 
nonzero drift, however small it may be. The detailed technical description
of generation of white noise (with zero and nonzero mean) from Brownian
motion has been included in Appendix.
Therefore, if $X(t)$ is the random displacement variable of a Brownian motion 
with drift coefficient $m$, its probability density function $P(X(t))$ can be written as 
 \begin{equation}
 P(X(t))=\frac{1}{\sqrt{2\pi t}\sigma}exp\left[-\frac{(X(t)-mt)^2}{2\sigma^2t}\right],
 \label{brownian0}
 \end{equation}
 where $\sigma\sqrt{t}$ is the standard deviation of the distribution and $t$ the time.
Taking the stochastic time derivative of $X(t)$, we obtain the white noise process which we denote by
$\eta(t)$~($=\dot{X}(t)$). Since the stochastic variable $X(t)$ is not differentiable in the usual sense,
we consider a finite difference
approximation of $\eta(t)$ using a time interval of width $\Delta t$ as
\begin{eqnarray}
\eta_{\Delta t}(t)=\frac{X(t+\Delta t)-X(t)}{\Delta t}.
\label{whitenoisesupp0}
\end{eqnarray}

Therefore, the presence of infinitesimal molecular viscosity (and shear), which is 
there always, would be 
enough just to give rise to a nonzero (infinitesimal) temperature, 
leading to thermal noise which can do the rest of the job of governing instability.
Note that a very tiny mean noise strength, due to tiny asymmetry in the system, 
is enough to lead to linear instability,
as demonstrated in previous sections. Here, the externally applied force (for plane Couette)
or the force arising due to the presence of strongly gravitating object (accretion disk)
introduces the asymmetry in the system, just like, e.g., the Brownian ratchets 
which has several applications in soft condensed matter and biology 
(see, e.g., \citealt{nzerodrft}). The measure of asymmetry and drag determines
the value of $m$, which furthermore controls the growth rate of perturbation.
The corresponding power spectrum appears to
be almost flat/constant (for ideal white noise it is purely flat).
Although in our chosen shearing box, the azimuthal direction is                 
assumed to be periodic, every such small box always encounters drag
and hence thermal fluctuation, which assures the presence of nonzero mean 
noise. As a result, every such sharing box reveals exponential growth
of perturbation. 


\section{Discussion and conclusions}\label{conclusion}

We have shown that linearly perturbed hydrodynamic apparently Rayleigh stable rotating shear flows, including accretion disks, 
and plane Couette flow, driven stochastically, can indeed 
be unstable, since the averaged
values of the perturbations grow exponentially. Due to background shear and hence drag, 
thermal fluctuations arise in these flows which induce Brownian motion 
of the fluid particles and hence stochastic forcing by white noise. Therefore the accretion flows, 
in particular due to perturbation, are inevitably 
driven by white noise which cannot be neglected. It is indeed shown in experiments that
the stochastic details decide whether turbulence will spread or eventually decay \citep{avilasci}, which furthermore argues
for the determining factor played by stochastic forcing, which we demonstrate here for the first time.
Since the forcing term in this system is a random variable, the solutions of the perturbations 
$u(\vec{x},t), \zeta(\vec{x},t)$ are also random variables and hence have some distributions whose averaged values are 
investigated.
Hence, we have shown that even in the absence of magnetic field, accretion disks 
can be made unstable and plausibly turbulent if they are driven by stochastic noise which is very likely to be 
present in the disks due to thermal fluctuations. In fact, we argue that neglecting the
stochastic noise in accretion flows and any other shear flows is 
vastly an inappropriate assumption. This is because, some shear is always 
there (because those are always driven externally by definition), which leads to some temperature (however be the magnitude) and
a small temperature is enough to reveal stochastic noise, which is the basic
building block of our work. Hence, the presence of (asymmetric) drag and stochastic noise
in shearing flows is inherent.
Hence, this work inevitably presents the origin of pure hydrodynamic instability of 
rotating shear flows and plane Couette flow. 
Therefore, this sheds enormous
light on to the understanding of formation of planets and stars.

Evidently this mechanism works for magnetized shear flows as well, because 
thermal fluctuations are available there also. For example, a background field 
of the order of unity with $m/A_0\sim 10^8$ can easily lead to unstable
modes of perturbation for $\alpha\gtrsim 5$ in the limit of very large $Re$ and 
$Rm$ which is the case in accretion disks. In future, we will report this result
in detail. Indeed, earlier we studied stochastically driven magnetized
flows and showed them to be plausibly unstable and turbulent by 
calculating the correlation functions of perturbations 
(\citealt{nath1,nath2}).
Hence the pure hydrodynamic instability explored here is generic. 
This is, to the best of our knowledge, the first solution
to the century old problem of hydrodynamic instability of apparently Rayleigh stable flows.
In due courses, one has to investigate how exactly
the required value of stochastic forcing strength could be arised in real
systems and if the growth rates of unstable modes
could adequately explain data. In certain cases, only high $Re$ reveals instability
which might be difficult to achieve in laboratory experiments and numerical
simulations as of now.

\appendix

\section{Generation of white noise from random walk via Brownian motion}\label{rw} 
We have assumed here that the white noise has a nonzero mean value. 

The term {\it white noise} is ambiguous. 
To shed light on this matter, here we point out the two definitions of white noise. \cite{brown} 
defines it as \\ \\ 
\textquotedblleft . . . a stationary random process having a constant spectral density function.\textquotedblright.\\ \\ 
\cite{pap} defines it as \\ \\ 
\textquotedblleft We shall say that a process $v(t)$ is white noise if its values $v(t_i)$ 
and $v(t_j)$ are uncorrelated for every $t_i$ and $t_j\neq t_i$: $C(t_i,t_j)=0, t_i\neq t_j$.\textquotedblright. \\ \\ 

The following subsections explore the implications of each definition with respect to the mean of the resulting 
process. 
\subsection{White noise as a stochastic process with constant power spectral density (Brown's definition)} 

Let $X(t)$ is an {\it ergodic stochastic process} with the property that it has a constant power spectral density, i.e. 
\begin{eqnarray}
\Phi_{xx}(\omega)=\alpha,
\label{psd}
\end{eqnarray}
where $\Phi_{xx}(\omega)$ is the power spectral density of the random variable $X(t)$ and $\alpha$ is a constant. 
Then the corresponding autocorrelation function for the process is 
\begin{eqnarray} 
E[X(t)X(t+\tau)]=\Phi_{xx}(\tau)=\alpha \delta(\tau),
\label{autocorr}
\end{eqnarray}
by taking inverse Fourier transform of $\Phi_{xx}(\omega)$,
where $E\left[\cdot\right]$ denotes the expectation value. Now let us 
assume that $X(t)$ is a zero mean white noise process and $Y(t)=X(t)+m$ is a nonzero mean process. Then 
\begin{eqnarray} 
\nonumber
\Phi_{yy}(\tau)=E[Y(t)Y(t+\tau)]=E[(X(t)+m)(X(t+\tau)+m)]=\alpha\delta(\tau)+m^2. \\ 
\end{eqnarray}
Therefore 
\begin{eqnarray}
\Phi_{yy}(\omega)=\alpha+2\pi m^2\delta(\omega),
\end{eqnarray}
which is not constant, thus $Y(t)$ violates the requirement of the white noise process by this definition.

\subsection{White noise as an uncorrelated stochastic process (Papoulis' definition)} 
Representing the Papoulis' definition of white noise in our notations, we can write, a stochastic process $X(t)$ 
is called a white noise process if any two distinct random variables of this stochastic process are independent 
and uncorrelated, i.e., the autocovariance function $C(X(t),X(t+\tau))=0$ when $\tau\neq 0$. In mathematical notation, 
\begin{eqnarray} 
\nonumber
C(X(t),X(t+\tau))=E[(X(t)-m_t)(X(t+\tau)-m_{t+\tau})] \\ 
\nonumber
=E[X(t)]E[X(t+\tau)]-m_tm_{t+\tau} \\ 
=m_tm_{t+\tau}-m_tm_{t+\tau}=0,
\label{cov}
\end{eqnarray}
where $m_t$ and $m_{t+\tau}$ are the corresponding mean values of the random variables $X(t)$ and $X(t+\tau)$ 
respectively. We can write the second equality in equation (\ref{cov}) since $X(t)$, for different values of $t$, 
are independent random variables by definition. 
{\it Thus it is not necessary that a white noise process always has to have a zero mean, from Papoulis' definition}. 
That is, a stochastic process having nonzero mean can be a white noise process according to this definition. 
{\it In the present work, we have used Papoulis' definition of white noise which can indeed have a non-zero mean}. Now let us explain why we 
have chosen Papoulis' definition over Brown's definition. 

\subsection{Why ideal white noise, having a constant power spectral density, is impossible in reality?}
Let us consider a signal $f(t)$ with constant power spectral density $\Phi_{ff}(\omega)$. That is, 
\begin{eqnarray}
\Phi_{ff}(\tau)=E[f(t)f(t+\tau)],
\label{corrf}
\end{eqnarray}
and the Fourier transform of $\Phi_{ff}(\tau)$ is $\Phi_{ff}(\omega)=$ a constant. Now the {\it Parseval's theorem} 
tells us that
\begin{eqnarray}
\int_{-\infty}^\infty |f(t)|^2dt=\int_{-\infty}^\infty |f(\omega)|^2d\omega, 
\label{energyf}
\end{eqnarray}
where $f(\omega)$ is the Fourier transform of $f(t)$. 
Since $\Phi_{ff}(\omega)$ and consequently $|f(\omega)|^2$ has a constant 
positive value according to Brown's definition of white noise, 
the equation (\ref{energyf}) tells us that the total power of the signal is infinity 
(see, e.g., \citealt{gardiner,kuo,poor,zhong,klie}). In mathematical terminology, 
the energy norm of the signal $f(t)$ is infinity and hence the function $f(t)$ is not $L^2$-integrable. Therefore, driving 
a system by a stochastic noise with constant power spectral density is same as injecting infinite amount of energy into 
the system, which is unphysical \citep{gardiner}. 

\subsection{Generation of Brownian motion (with zero and nonzero drift) from random walk} 
In this section, we outline 
the derivation of white noise starting from the random walk, via Brownian motion. 
%
\begin{figure}[tbp]
\centering
\includegraphics[angle=0,height=1.0cm,width=9.0cm]{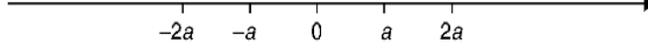}
\caption{A diagram of an one dimensional random walk.}
\label{rwdiagram}
\end{figure} 
\vskip1.5cm
Figure \ref{rwdiagram} shows an array of positions $ja$ where $j=0, \pm1, \pm2,$ etc. and $a$ is the spacing 
between points. At each interval of time, $\tau$, a hop is made with probability $p$ to the right and $q=1-p$ 
to the left. The distribution of $r$, of hops to the right, in $N$ steps is given by the Bernoulli distribution
\begin{eqnarray}
P_N(r)=\frac{N!}{r!(N-r)!}p^rq^{N-r}.
\label{bar}
\end{eqnarray}
The first moment (mean) and the second moment (variance) of the Bernoulli distribution in equation (\ref{bar}) is 
given by 
\begin{eqnarray} 
\nonumber
\langle r\rangle=Np, \\ 
\langle (\Delta r)^2\rangle=Npq.
\label{meanvar}
\end{eqnarray}
A particle that started at $0$ and took $r$ steps to the right and $N-r$ steps to the left arrives at the position 
\begin{eqnarray} 
n=r-(N-r)=2r-N, 
\label{position}
\end{eqnarray} 
with mean value 
\begin{eqnarray} 
\langle n\rangle=N(2p-1)=N(p-q).
\label{meanposition}
\end{eqnarray}
Notice that, if $p=q=1/2$, or equal probability to jump to the right and the left, the average position 
after $N$ steps will remain $0$. The second moment about the mean is given by 
\begin{eqnarray} 
\langle (\Delta n)^2\rangle=4\langle (\Delta r)^2\rangle=4Npq. 
\label{varposition}
\end{eqnarray}
Therefore, from the central limit theorem, the limiting distribution after many steps is Gaussian, with 
the first and second moments just obtained (in equations (\ref{meanposition}) and (\ref{varposition})), given by 
\begin{eqnarray} 
P_N(n)=\frac{1}{\sqrt{2\pi}\sqrt{(4Npq)}}~exp\left\lbrace\frac{-\left[n-N(p-q)\right]^2}{8Npq}\right\rbrace. 
\label{distposition}
\end{eqnarray}
If we introduce the position and time variables by the relations 
\begin{eqnarray} 
\nonumber 
x=na, \\  
N=t/\tau,
\label{newtimepos}
\end{eqnarray}
the moments of $x$ are given by 
\begin{eqnarray} 
\nonumber
\langle x\rangle=N(p-q)a=(p-q)at/\tau, \\ 
\langle (\Delta x)^2\rangle=4Npqa^2=\left(\frac{4pqa^2}{\tau}\right)t=2Dt.
\label{meanvarnew}
\end{eqnarray}
The factor $2$ in the definition of diffusion coefficient $D$ is appropriate for one dimension, and would 
be replaced by $2d$ if we consider the random walk in a space of dimension $d$. Thus the distribution moves 
with a \textquotedblleft drift\textquotedblright velocity 
\begin{eqnarray} 
v=(p-q)a/\tau, 
\label{driftvel}
\end{eqnarray}
and spreads with a diffusion coefficient defined by 
\begin{eqnarray} 
D=\frac{2pqa^2}{\tau}.
\label{diffcoeff}
\end{eqnarray}
Thus the probability distribution of the displacement $x$ of a particle under this random walk is 
\begin{eqnarray} 
P(x)=\frac{1}{\sqrt{2\pi t}\sigma}~exp\left[-\frac{(x-vt)^2}{2\sigma^2t}\right],
\label{distdisp}
\end{eqnarray}
where $\sigma=\sqrt{2D}$. A stochastic process in which the random variables $X(t)$s are stationary and 
independent and have distribution as in equation (\ref{distdisp}), is called a Brownian motion or Wiener process. 
It is very clear from the equation (\ref{driftvel}) that, when $p=q=1/2$, then the drift velocity is $0$, which means 
if some random walk is fully symmetric without any bias, then only we obtain the zero drift velocity of the corresponding 
Brownian motion (which is known as standard Brownian motion in literature). However, if some process has 
any asymmetry (for example hydrodynamic flows with shear in a particular direction, bulk hydrodynamic flows, 
flows under gravity etc.), the random walk of particles in that process will have some bias (i.e. $p\neq q$), 
which eventually introduces a Brownian motion with {\it nonzero drift velocity}. 

\subsection{White noise from Brownian motion} 
If we take stochastic time derivative of a Brownian motion or Wiener process, we obtain a {\it white noise} 
process. If $X(t)$ is the random displacement variable of a Brownian motion 
with drift velocity $m$, its probability density function $P(X(t))$ can be written as (using equation (\ref{distdisp})) 
\begin{eqnarray}
P(X(t))=\frac{1}{\sqrt{2\pi t}\sigma}exp\left[-\frac{(X(t)-mt)^2}{2\sigma^2t}\right],
\label{browniansupp}
\end{eqnarray}
where $\sigma\sqrt{t}$ is the standard deviation of the distribution and $t$ the time.
Taking the stochastic time derivative of $X(t)$, we obtain the white noise process which we denote by 
$\eta(t)$~($=\dot{X}(t)$). Since the stochastic variable $X(t)$ is not differentiable in the usual sense, 
we consider a finite difference 
approximation of $\eta(t)$ using a time interval of width $\Delta t$ as 
\begin{eqnarray}
\eta_{\Delta t}(t)=\frac{X(t+\Delta t)-X(t)}{\Delta t}.
\label{whitenoisesupp}
\end{eqnarray}
Since the stochastic random variables $X(t)$ corresponding to a Brownian motion process are stationary and independent, 
from equations (\ref{browniansupp}) and (\ref{whitenoisesupp}) we obtain that the white noise process has mean/averaged value $m$ 
and variance $\left(\sigma^2/\Delta t+2\sigma^2t/\Delta t^2\right)$. As $\Delta t\to 0$, the variance 
$\left(\sigma^2/\Delta t+2\sigma^2t/\Delta t^2\right)\to\infty$, and this white 
noise tends to the ideal white noise having a constant power spectral density \citep{kuo}. However, since Brownian motion is 
not differentiable anywhere, the ideal white noise does not exist, as also explained above from the energy norm 
point of view. 

Now we will show that the white noise defined in equation (\ref{whitenoisesupp}) satisfies the Papoulis' definition 
of white noise, i.e., the process is an uncorrelated stochastic process. To establish this, let us first note 
that if $X(t)$ and $X(s)$ are two random variables of a Brownian motion with $s\leq t$, then 
\begin{eqnarray} 
\nonumber
C(X(t),X(s))=E[(X(t)-mt)(X(s)-ms)] \\ 
\nonumber
=E[\lbrace(X(t)-mt)-(X(s)-ms)+(X(s)-ms)\rbrace (X(s)-ms)] \\ 
\nonumber
=E[\lbrace(X(t)-X(s))-(mt-ms)\rbrace(X(s)-ms)]+E[(X(s)-ms)^2] \\ 
=0+\sigma^2 s=\sigma^2 {\rm min}\lbrace t,s\rbrace.
\label{browncov}
\end{eqnarray}
The third equality is possible since $(X(t)-X(s))$ and $X(s)$ are independent random variables for a Brownian motion.
Having the result of equation (\ref{browncov}) in hand, we now calculate the autocovariance of white noise. It is very 
easy to verify that the autocovariance function $C(X,Y)$ of two random variables $X$ and $Y$ is a linear function 
in both of its arguments. Therefore, 
\begin{eqnarray} 
\nonumber
C(\eta_{\Delta t}(t),\eta_{\Delta t}(s))=C\left[\frac{X(t+\Delta t)-X(t)}{\Delta t},\frac{X(s+\Delta t)-X(s)}{\Delta t}\right] \\ 
\nonumber
=\frac{1}{\Delta t^2} \left[C(X(t+\Delta t),X(s+\Delta t))-C(X(t+\Delta t),X(s)) \right. \\ 
\left. -C(X(t),X(s+\Delta t))+C(X(t),X(s))\right].
\label{whitecov}
\end{eqnarray}
When $|t-s|\leq\Delta t$, i.e. $s-\Delta t\leq t\leq s+\Delta t$, then using equation (\ref{browncov}), 
from equation (\ref{whitecov}) we obtain 
\begin{eqnarray} 
\nonumber
C(\eta_{\Delta t}(t),\eta_{\Delta t}(s))=\sigma^2\frac{1}{\Delta t^2} ({\rm min}\lbrace t,s\rbrace+\Delta t-s-t+{\rm min}\lbrace t,s\rbrace) \\  
\nonumber 
=\sigma^2\frac{1}{\Delta t}\left\lbrace 1-\left(\frac{s+t-2{\rm min}\lbrace t,s\rbrace}{\Delta t}\right)\right\rbrace \\ 
=\sigma^2\frac{1}{\Delta t}\left( 1-\frac{|t-s|}{\Delta t}\right).
\label{whitecovmid}
\end{eqnarray}
Now let us consider the cases when $|t-s|\geq\Delta t$, i.e. when $s+\Delta t\leq t$ or $t+\Delta t\leq s$. For 
$s+\Delta t\leq t$, equations (\ref{browncov}) and (\ref{whitecov}) imply
\begin{eqnarray} 
C(\eta_{\Delta t}(t),\eta_{\Delta t}(s))=\sigma^2\frac{1}{\Delta t^2} (s+\Delta t-s-(s+\Delta t)+s)=0, 
\label{whitecovleft}
\end{eqnarray}
and also for $t+\Delta t\leq s$, 
\begin{eqnarray} 
C(\eta_{\Delta t}(t),\eta_{\Delta t}(s))=\sigma^2\frac{1}{\Delta t^2} (t+\Delta t-(t+\Delta t)-t+t)=0. 
\label{whitecovright}
\end{eqnarray}
Therefore, 
\begin{eqnarray} 
\nonumber
C(\eta_{\Delta t}(t),\eta_{\Delta t}(s))=\sigma^2\frac{1}{\Delta t}\left( 1-\frac{|t-s|}{\Delta t}\right), 
~{\rm when}~|t-s|\leq\Delta t \\ 
=0, {\rm otherwise},
\label{whitecovfinal}
\end{eqnarray}
i.e. $\eta_{\Delta t}(t)$ and $\eta_{\Delta t}(s)$ are uncorrelated. Let us define 
\begin{eqnarray} 
\nonumber
t-s=\tau,~{\rm and},~\delta_{\Delta t}(\tau)=\frac{1}{\Delta t}\left( 1-\frac{|\tau|}{\Delta t}\right), 
~{\rm when}~|\tau|\leq\Delta t \\ 
=0, {\rm otherwise}.
\label{whitecovdelta}
\end{eqnarray}
%
%
\begin{figure}[tbp]
\centering
  \begin{tabular}{@{}cc@{}}
    \includegraphics[width=.55\textwidth]{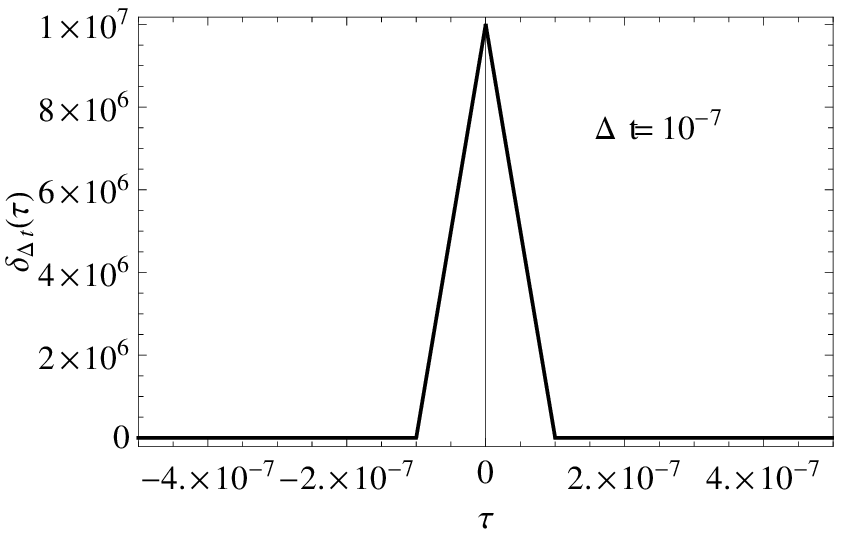} \\ \\
    \includegraphics[width=.6\textwidth]{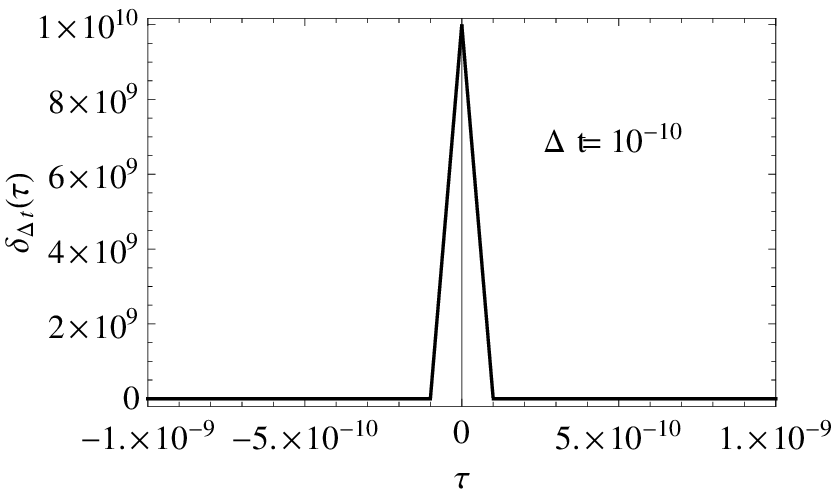} \\ \\ 
    \end{tabular}
  \caption{Variation of $\delta_{\Delta t}(\tau)$ for two small values of $\Delta t$.}
\label{twodelta}
\end{figure} 
Figure \ref{twodelta} shows the variation of $\delta_{\Delta t}(\tau)$ for two different small values of $\Delta t$. The function 
$\delta_{\Delta t}(\tau)$ defined in equation (\ref{whitecovdelta}) is an approximation of the well known delta function 
$\delta(\tau)$, because 
\begin{eqnarray} 
\nonumber
\lim_{\Delta t\to 0}\delta_{\Delta t}(\tau)=\delta(\tau),
\label{whitecovdeltalimit}
\end{eqnarray}
as it is seen from Fig. 2. Also the function $\delta_{\Delta t}(\tau)$ satisfies the integral property of the delta function 
as shown below, 
\begin{eqnarray} 
\nonumber
\int_{-\infty}^\infty\delta_{\Delta t}(\tau)d\tau=\int_{-\infty}^\infty\frac{1}{\Delta t}\left(1-\frac{|\tau|}{\Delta t}\right)d\tau \\ 
=\frac{1}{\Delta t}\int_{-\Delta t}^0\left(1+\frac{\tau}{\Delta t}\right)d\tau
+\frac{1}{\Delta t}\int_0^{\Delta t}\left(1-\frac{\tau}{\Delta t}\right)d\tau=1.
\label{whitedeltaint}
\end{eqnarray}
Hence, when $\Delta t\to 0$, from equation (\ref{whitecovfinal}) we obtain 
\begin{eqnarray} 
\nonumber
C(\eta_{\Delta t}(t),\eta_{\Delta t}(s))=\sigma^2\delta(\tau)=\sigma^2\delta(t-s). 
\label{whitecovuncorr}
\end{eqnarray}
Therefore, the noise with nonzero mean, obtained from the stochastic time derivative of Brownian motion with nonzero drift, 
is a white noise process according to the Papoulis' definition and also has the correlators as defined below 
equation (\ref{hydroorrzeta}). 


\section*{Acknowledgments}

The authors acknowledge partial support through research Grant No. ISTC/PPH/BMP/0362.
The authors thank Amit Bhattacharjee of IISc and Debashish Chowdhury of IIT Kanpur
for discussions related to possible nonzero mean of white noise and the Brownian
ratchets. Thanks are also due to the anonymous referee and Ramesh Narayan of Harvard 
for suggestions which have helped to improve
the presentation of the paper.



\begin{thebibliography}{30}
\expandafter\ifx\csname
natexlab\endcsname\relax\def\natexlab#1{#1}\fi

\bibitem[Afshordi et al. (2005)]{amn} Afshordi, N., Mukhopadhyay, B.,~\& Narayan, R. 2005, \apj, 629, 373 
\bibitem[Avila (2012)]{avila} Avila, M. 2012, Phys. Rev. Lett., 108, 124501 
\bibitem[Avila et al. (2011)]{avilasci} Avila, M. {\it et al.} 2011, Science, 333, 192  
\bibitem[Balbus (2011)]{balbusnature} Balbus, S.~A. 2011, Nature, 470, 475 
\bibitem[Balbus \& Hawley (1991)]{bh} Balbus, S.~A.,~\& Hawley, J.~F. 1991, \apj, 376, 214 
\bibitem[Stone et al. (1996)]{balbusetal96} Balbus, S.~A., Hawley, J.~F.,~\& Stone, J.~M. 1996, \apj, 467, 76 
\bibitem[Barker \& Latter (2015)]{barker} Barker, A. J., \& Latter, H. N. 2015, MNRAS, 450, 21
\bibitem[Barkley et al. (2015)]{fulltur} Barkley D. {\it et al.} 2015, Nature, 526, 550 
\bibitem[Bottin \& Chat\'e (1998)]{stattur} Bottin, S.,~\& Chat\'e, H. 1998, Eur. Phys. J. B, 6, 143 
\bibitem[Brown (1983)]{brown} Brown, R.~G. 1983,~{\it Introduction to Random Signal Analysis and Kalman Filtering.}, John Wiley and Sons 
\bibitem[Cantwell et al. (2010)]{berk} Cantwell, C.~D., Barkley, D.,~\& Blackburn, H.~M. 2010, Phys. Flud., 22, 034101 
\bibitem[Chandrasekhar (1960)]{chandra} Chandrasekhar, S. 1960, Proc. Nat. Acad. Sci., 46, 53 
\bibitem[Dauchot \& Daviaud (1995)]{dau} Dauchot, O.,~\& Daviaud, F. 1995, Phys. Fluids, 7, 335 
\bibitem[De Dominicis \& Martin (1979)]{dm} De Dominicis, C.,~\& Martin, P.~C. 1979, Phys. Rev. A, 19, 419 
\bibitem[Dubrulle et al. (2005)]{dub1} Dubrulle, B., Dauchot, O., Daviaud, F., Longaretti, P.~-Y., Richard, D.,~\& 
Zahn, J.~-P. 2005, Phys. Fluids, 17, 095103 
\bibitem[Marie et al. (2005)]{dub2} Dubrulle, B., Marie, L., Normand, C., Hersant, F., Richard, D.,~\& Zahn, J.~-P. 2005, 
A\&A, 429, 1 
\bibitem[Forster et al. (1977)]{nelson} Forster, D., Nelson, D.~R.,~\& Stephen, M.~J. 1977, 
Phys. Rev. A, 16, 732 
\bibitem[Fromang \& Papaloizou (2007)]{fromang-papaloizou07} Fromang, S.,~\& Papaloizou, J. 2007, A\&A, 476, 1113 
\bibitem[Gardiner (1985)]{gardiner} Gardiner, C.~W. 1985,~{\it Handbook of Stochastic Methods for Physics, Chemistry and Natural Sciences}, 
2nd edition, Springer  
\bibitem[Gu et al. (2000)]{gu} Gu, P.~-G., Vishniac, E.~T.,~\& Cannizzo, J.~K. 2000, ApJ, 534, 380 
\bibitem[Hawley et al. (1999)]{hawleyetal99} Hawley, J.~F., Balbus, S.~A.,~\& Winters, W.~F. 1999, \apj, 518, 394 
\bibitem[Kile (1995)]{klie} Kliemann, W.,~\& Namachchivaya, S. 1995,~{\it Nonlinear Dynamics and Stochastic Mechanics}, 
CRC Press 
\bibitem[Kim \& Ostriker (2000)]{kim} Kim, W.~-T.,~\& Ostriker, E.~C. 2000, ApJ, 540, 372 
\bibitem[Klahr \& Bodenheimer (2003)]{klar1} Klahr, H.~H.,~\& Bodenheimer, P. 2003, \apj, 582, 869 
\bibitem[Klahr \& Hubbard (2014)]{klar} Klahr, H.,~\& Hubbard, A. 2014, 
\apj, 788, 21
\bibitem[Kuo (1996)]{kuo} Kuo, H.~H. 1996,~{\it White Noise Distribution Theory}, 
CRC Press 
\bibitem[Latter (2016)]{latter} Latter, H.~N. 2016, MNRAS, 455, 2608
\bibitem[Lesur \& Longaretti (2005)]{ll05} Lesur, G.,~\& Longaretti, P.~-Y. 2005, A\&A, 444, 25 
\bibitem[Lin \& Youdin (2015)]{lin} Lin, M.-K., \& Youdin, A. N. 2015, \apj, 811, 17
\bibitem[Luki\'c et al. (2005)]{prlf} Luki\'c, B., Jeney, S., Tischer, C., Kulik, A.~J., Forr\'o, L.,~\& 
Florin, E.-L. 2005, Phys. Rev. Lett., 95, 160601 
\bibitem[Lyra (2014)]{lyra} Lyra, W. 2014, \apj, 789, 77
\bibitem[Mahajan \& Krishan (2008)]{mk} Mahajan, S.~M.,~\& Krishan, V. 2008, \apj, 682, 602 
\bibitem[Marcus et al. (2015)]{marcus1} Marcus, P.~S., Pei, S., Jiang, 
C.-H.,~\& Barranco, J.~A. 2015, \apj, 808, 87
\bibitem[Marcus et al. (2013)]{marcus2} Marcus, P.~S., Pei, S., Jiang, 
C.-H.,~\& Hassanzadeh, P. 2013, Phys. Rev. Lett., 111, 084501
\bibitem[Miyazaki \& Bedeaux (1995)]{physica} Miyazaki, K.,~\& Bedeaux, D. 1995, Physica A, 217, 53 
\bibitem[Mukhopadhyay (2013)]{bmplb} Mukhopadhyay, B. 2013, Phys. Lett. B, 721, 151 
\bibitem[Mukhopadhyay et al. (2005)]{man} Mukhopadhyay, B., Afshordi, N.,~\& Narayan, R. 2005, \apj, 629, 383 
\bibitem[Mukhopadhyay \& Chattopadhyay (2013)]{mc13} Mukhopadhyay, B.,~\& Chattopadhyay, A.~K. 2013, J. Phys. A, 46, 035501 
\bibitem[Mukhopadhyay et al. (2011)]{bmraha} Mukhopadhyay, B., Mathew, R.,~\& 
Raha, S. 2011, 
NJPh, 13, 023029 
\bibitem[Nath \& Chattopadhyay (2014)]{nath2} Nath, S.~K.,~\& 
Chattopadhyay~A.~.K.~2014, Phys. Rev. E, 90, 063014
\bibitem[Nath \& Mukhopadhyay (2015)]{tran} Nath, S.~K.,~\& Mukhopadhyay, B. 2015, Phys. Rev. E, 92, 023005 
\bibitem[Nath et al. (2013)]{nath1} Nath, S.~K.,~Mukhopadhyay, B.,~\&
Chattopadhyay~A.~.K.~2013, Phys. Rev. E, 88, 013010
\bibitem[Nelson et al. (2013)]{nelson} Nelson, R.~P., Gressel, O., 
\& Umurhan, O.~M. 2013, MNRAS, 435, 2610
\bibitem[Paoletti et al. (2012)]{paoletti} Paoletti, M.~S., van Gils, D.~P.~M., Dubrulle, B., Sun, C., Lohse, D.,~\& 
 Lathrop, D.~P. 2012, A\&A, 547, A64 
\bibitem[Papoulis (1991)]{pap} Papoulis, A. 1991,~{\it Probability, Random Variables, and Stochastic Processes, 3rd ed.}, WCB/McGraw-Hill 
\bibitem[Poor (2013)]{poor} Poor, H.~V. 2013,~{\it An Introduction to Signal Detection and Estimation}, 
Springer Science \& Business Media 
\bibitem[Pringle (1981)]{pringle}  Pringle, J.~E. 1981, \araa, 19, 137 
\bibitem[Pumir (1996)]{pumir96} Pumir, A. 1996, Phys. Fluids, 8, 3112 
\bibitem[Richard et al. (2016)]{ric} Richard, S., Nelson, R. P., \& Umurhan, O. M. 2016, MNRAS, 456, 3571
\bibitem[Richard \& Zahn (1999)]{zahn} Richard, D.,~\& Zahn, J.~-P. 1999, A\&A, 347, 734 
\bibitem[Rudiger \& Zhang (2001)]{rud} Rudiger, G.,~\& Zhang, Y. 2001, A\&A, 378, 302 
\bibitem[Shakura \& Sunyaev (1973)]{ss73} Shakura, N. I., \& Sunyaev, R. A. 1973, A\&A, 24, 337
\bibitem[Stoll \& Kley (2014)]{stoll} Stoll, M.~H.~R., \& Kley, W. 2014, A\&A, 572, A77
\bibitem[Stoll \& Kley (2016)]{stoll2} Stoll, M.~H.~R., \& Kley, W. 2016, arXiv:1607.02322
\bibitem[Trefethen et al. (1993)]{tref} Trefethen, L.~N., Trefethen, A.~E., Reddy, S.~C.,~\& Driscoll, T.~A. 1993, 
Science, 261, 578 
\bibitem[Oudenaarden \& Boxer (1999)]{nzerodrft} van Oudenaarden, A.,~\& Boxer, S.~G. 1999, Science, 85, 1046 
\bibitem[Umurhan et al. (2016)]{um} Umurhan, O.~M., Nelson, R.~P.,~\& 
Gressel, O. 2016, A\&A, 586, A33
\bibitem[Velikhov (1959)]{velikhov} Velikhov, E. 1959, J. Exp. Theor. Phys., 36, 1398
\bibitem[Yecko (2004)]{yecko} Yecko, P.~A. 2004, \aap, 425, 385 
\bibitem[Zhong (2006)]{zhong} Zhong, W.~X. 2006,~{\it Duality System in Applied Mechanics and Optimal Control}, 
Springer Science \& Business Media  


\end{thebibliography}
\end{document}